\documentclass[12pt,preprint]{aastex}
\usepackage{graphicx,float,subfigure,amsmath,amsthm,amssymb,multirow}
\newcommand{\etal}{et~al.\ }

\newcommand{\cmsq}{\hbox{cm$^{-2}$}}

\newcommand{\flux}{\hbox{erg~cm$^{-2}$~s$^{-1}$}}

\newcommand{\lumin}{\hbox{erg~s$^{-1}$}}
\newcommand{\lumind}{\hbox{erg~s$^{-1}$~Hz$^{-1}$}}

\newcommand{\aox}{$\alpha_{\rm ox}$}
\newcommand{\nh}{\hbox{${N}_{\rm H}$}}

\newcommand{\be}{\begin{equation}}
\newcommand{\ee}{\end{equation}}
\newcommand{\ba}{\begin{eqnarray}}
\newcommand{\ea}{\end{eqnarray}}

\newcommand{\chandra}{\emph{Chandra}}

\newcommand{\suz}{\emph{Suzaku}}
\newcommand{\mass}{\emph{2MASS}}
\newcommand{\sdss}{\emph{Sloan Digital Sky Survey}}

\newcommand{\simgt}{\lower 2pt \hbox{$\, \buildrel {\scriptstyle >}\over {\scriptstyle\sim}\,$}}
\newcommand{\simlt}{\lower 2pt \hbox{$\, \buildrel {\scriptstyle <}\over {\scriptstyle\sim}\,$}}
\newcommand{\ls}{\lower 2pt \hbox{$\;\scriptscriptstyle \buildrel<\over\sim\;$}}
\newcommand{\gs}{\lower 2pt \hbox{$\;\scriptscriptstyle \buildrel>\over\sim\;$}}

\begin{document}

\def\arcsec{$^{\prime\prime}$}
\def\arcmin{$^{\prime}$}
\def\degr{$^{\circ}$}

\title{Suzaku Observations of Three FeLoBAL QSOs, SDSS J0943+5417, J1352+4239, and J1723+5553}

\author{Leah K. Morabito\altaffilmark{1}, Xinyu Dai\altaffilmark{1}, Karen M. Leighly\altaffilmark{1}, Gregory R. Sivakoff\altaffilmark{2}, Francesco Shankar\altaffilmark{3}}
\altaffiltext{1}{Homer L. Dodge Department of Physics \& Astronomy,
University of Oklahoma, Norman, OK 73019, USA, morabito@nhn.ou.edu}
\altaffiltext{2}{Department of Astronomy, University of
Virginia, Charlottesville, VA 22904, USA}
\altaffiltext{3}{Max-Planck-Instit\"{u}t f\"{u}r Astrophysik,
Karl-Schwarzschild-Str. 1, D-85748, Garching, Germany}

\begin{abstract}
We present \suz\ observations of three iron low-ionization broad absorption line quasars (FeLoBALs).
We detect J1723+5553 ($3\sigma$) in the observed 2--10 keV band, and constrain its intrinsic \nh\ column density to $\nh > 6\times10^{23}$ \cmsq\ 
by modeling its X-ray hardness ratio. We study the broadband spectral index, \aox, between the X-ray and UV bands by combining the X-ray
measurements and the UV flux extrapolated from \mass\ magnitudes, assuming a range of intrinsic column densities, and then comparing
the \aox\ values for the three FeLoBALs with those from a large sample of normal quasars. We find that the FeLoBALs are consistent
with the spectral energy distribution (SED) of normal quasars if the intrinsic \nh\ column densities are $\nh > 7\times10^{23}$ \cmsq\ 
for J0943+5417, $\nh > 2\times10^{24}$ \cmsq\ for J1352+4293, and $6\times10^{23} < \nh < 3\times10^{24}$ \cmsq\ for J1723+5553.
At these large intrinsic column densities, the optical depth from Thomson scattering can reach $\sim6$, which will significantly
modulate the UV flux. Our results suggest that the X-ray absorbing material could be located at a different place from the UV absorbing wind,
likely between the X-ray and UV emitting regions.
We find a significant kinetic feedback efficiency for FeLoBALs, indicating that the outflows are an important feedback mechanism in quasars.
\end{abstract}


\section{Introduction}
Quasars and active galactic nuclei (AGN) are ubiquitous phenomena wherever we look in the universe. The subset of quasars known as broad 
absorption line quasars (BALQSOs) are characterized by blue-shifted, rest-frame UV absorption troughs due to gas outflow. Even before the 
first definitive survey in \citet{b13}, BALQSOs provided a unique glimpse into the central engines of AGN, where the nature of the 
intrinsic absorption has important implications for the structure of the central engine. To classify a quasar as a BALQSO, the traditional definition of \citet{b13} 
requires the absorption troughs to be at least 2,000 km s$^{-1}$ wide. Recent studies have relaxed this, e.g., \citet{b14}, to include 
quasars with absorption troughs over 1,000 km s$^{-1}$ in width.  Regardless of which definition is used, BALQSOs are further divided 
into those that exhibit absorption troughs from only high-ionization species (HiBALs) and those with low-ionization species (LoBALs). 
Most BALQSOs are HiBALs, and most LoBALs also have high-ionization troughs in their spectra \citep{b13, b14}. LoBALs are further 
classified by the low-ionization species they exhibit: in this paper, the objects surveyed have strong iron absorption, and are 
therefore iron LoBALs (FeLoBALs). This type of BALQSO is rare, and comprises only 
1.5--2.1\% of the entire quasar population \citep{b12}, 
while LoBALs and BALQSOs make up $\sim$4--7\% \citep{b12} and $\sim$20--40\% \citep{b2,s08,gb08,k08,m08,a10}, respectively, depending on 
the threshold of the trough width. Focusing on FeLoBALs is meaningful because of their unique characteristics and the bearing they have 
on the broader picture: either as an evolutionary link in the normal AGN lifetime, or as a geometric interpretation of a uniform class of objects.

Besides the low-ionization absorption lines, the optical continua for LoBALs are more reddened than HiBALs and normal QSOs, suggestive 
of stronger dust absorption, perhaps from large quantities of dust in the vicinity of the central engine \citep{b15}. LoBALs in particular 
are more apparently X-ray weak than normal BALQSOs \citep{b16, b17}. There is a large portion, $\sim$80\%, not detected in X-rays, along with 
lower than expected values for the UV to X-ray luminosity ratio, \aox\footnote{The broadband spectral index, 2500\AA\ to 2 keV 
point-to-point power-law slope.} (\aox\ lower than expected by $\sim-0.9$), that leads most previous studies, e.g. 
\citet{b16,b17}, to conclude that there is extreme X-ray absorption in LoBALs. \citet{b23}, however, found that LoBALs in general 
have lower column densities ($N_H<10^{22}$ \cmsq) than HiBALs. This conclusion 
may have been influenced by their X-ray bright sample, required for sufficient quality to perform X-ray spectral analyses.  

Low-ionization BALQSOs with broad iron absorption lines (FeLoBALs) represent an extreme category. They have the reddest continuum spectra,
complex UV spectra, and possibly largest X-ray absorption column densities. There are a handful of FeLoBALs already studied in X-rays,
with only a few detections. Almost all were observed using the sensitive X-ray telescope, \chandra. The nearby FeLoBAL Mrk 231 
($z=0.042$) was also detected by \emph{XMM-Newton} and \emph{BeppoSAX} \citep{tk03,br04} in addition to \chandra\ \citep{g02b}, 
which allowed for the most in-depth studies of an FeLoBAL to date. All three studies find the 0.4--10 keV X-ray emission to be mostly reflected and 
scattered into our line of sight, and the direct X-ray emission is mostly absorbed. Using the PDS detector on board \emph{BeppoSAX}, 
\citet{br04} detect the direct X-ray emission of Mrk~231 at $3\sigma$ in the 15--60 keV band, which allows the authors to constrain an 
intrinsic column density $N_H\sim2\times10^{24}$ \cmsq. Another recent study focused on two FeLoBALs is \citet{b21}, where the authors 
calculated \aox\ upper limits for both objects, comparing them with a large sample of normal AGN \citep{b6}. Other studies focused on subsets 
of quasars, but had FeLoBALs contained within their scope. Quite a few of these focused on radio-detected objects. In one study, \citet{u05} 
selected 12 luminous red quasars, with the criteria that they were also detected by FIRST \citep{b95}, and used standard aperture photometry (for details see \S~\ref{sdr}) 
to conduct X-ray analyses. They found that all of their objects are X-ray absorbed to some degree; 
in addition, they found a slightly steeper spectral slope than that for normal quasars, with an unweighted mean of $\Gamma=2.2\pm0.4$ 
and a high-energy region weighted mean of $\Gamma=2.1\pm0.5$. \citet{br05} selected five radio-loud BALQSOs, three of which are LoBALs, 
and two of those LoBALs exhibited iron features. The three LoBALs were found to exhibit the steepest \aox\ values in the study. \citet{m09} also selected 21 
radio-loud BALQSOs, one of which is classified as an FeLoBAL. The authors proposed that the X-ray properties are explained by a model 
that consists of X-ray emission in the disk/corona and linked to the radio-jet. Almost all of these studies calculated intrinsic column 
density for these objects, most of which are in the 10$^{22}$ -- 10$^{24}$ \cmsq\ range. These will be discussed in detail in \S~\ref{tci}. In 
this study, we will use our X-ray detections and non-detections to constrain X-ray absorption column densities. One of our objects 
is the first X-ray detection (at the 3$\sigma$ level) by \suz\ of an FeLoBAL.

It is not understood why FeLoBALs possess unique physical characteristics; evolution and geometry are the contending theories. A geometric 
interpretation \citep[e.g.,][]{b18} proposes that the broad absorption line region of a BALQSO simply arises from looking straight down a 
narrow outflow, while normal AGN and quasars are the same object viewed from different angles. A recent discovery of a LoBAL transitioning 
from an FeLoBAL within a few years rest frame favors special lines of sight, because of the short time scale \citep{h10}. On the other hand, \citet{b20} proposed 
that BALQSOs, and FeLoBALs especially, are early progenitors of normal AGN and quasars, at the very end of an extreme starburst period and 
in the process of blowing out material from type II SNe that is Fe II rich. This evolutionary theory is partially supported 
by the higher fraction of LoBALs at high IR luminosities \citep[e.g.,][]{f07,u09,b12}. A recent \emph{Spitzer} spectral survey of six 
FeLoBALs \citep{f10} showed significant signatures of dust and PAH emissions. Spectral modelings in the UV regime also support a large 
covering fraction of the wind for FeLoBALs \citep[e.g.,][]{ca08}. Other evidence, such as the decreasing fraction as a function 
of radio luminosity, is consistent with a geometric interpretation \citep{s08,b12}. Therefore, the population of LoBALs and FeLoBALs could also be a 
combination of both evolution and geometry \citep{b12}. 

Another interesting question is how FeLoBALs contribute to the quasar feedback process. AGN feedback is widely used in galaxy formation models 
(e.g., Granato et al.\ 2004; Scannapieco \& Oh 2004; Hopkins et al.\ 2005; Shankar et al.\ 2008b) to explain the co-evolution of AGN and their 
host galaxies. Feedback can reproduce such phenomena as the $M$--$\sigma$ relation \citep[e.g.,][]{g11,g09}, star formation rates \citep{sn10}, 
and even the shortfall in the halo baryon fraction \citep{sn10}. Whether FeLoBALs are an evolutionary stage or a geometric interpretation of 
AGN, studying their feedback efficiency allows us to add to our limited picture of their physical characteristics.

In this paper we present new \suz\ observations of three FeLoBALs, SDSS objects J094317.59+541705.1, J135246.37+423923.5, and 
J172341.08+555340.5 (hereafter J0943+5417, J1352+4239, and J1723+5553, respectively). The targets are classified as FeLoBALs by \citet{b14}, 
and are the brightest ($K_s <$ 14.4 mag) of the \textit{2 Micron All Sky Survey} (\mass) selected BALQSOs \citep{b2}. J0943+5417 is 
a radio-intermediate quasar detected with a peak flux of 1.71 mJy/beam and RMS noise of 0.145 mJy/beam by Faint Images of the Radio Sky at Twenty-Centimeters 
(FIRST). None of these objects are detected at the flux limit in the NRAO/VLA Sky Survey \citep{c98}. 
J0943+5317 and J1352+4239 are typical FeLoBALs with extremely complex rest frame UV spectra, whereas J0943+5417 has absorption 
troughs so wide that they overlap and completely suppress the UV continuum. We extrapolate our UV data from \mass, 
which samples the rest frame optical bands, and then correct to the rest frame UV.  The complexity of the observed optical/rest frame UV spectra 
makes it extremely difficult to model the continuum.  This is addressed in \S~3. Since the targets all have redshifts $z\sim2$, the XIS 
on-board \suz\ can probe the rest frame energy up to 30~keV, where the rest frame 6--30 keV (observed 2--10 keV) emission is only moderately affected by absorption 
even if $\nh \sim 10^{24}\cmsq$, allowing us to detect the FeLoBALs if they are Compton thin. We study the relationship between X-ray and 
UV luminosities for these three FeLoBALs, and compare our values to a large sample of optically selected normal quasars. Our \suz\ observations 
and data reduction are described in \S \ref{xray}, followed by calculations of 2500\AA\ luminosities in \S 3. Section 4 presents results, 
followed by discussion in \S 5.  Throughout the paper we assume a cosmology of $H_0$=70 km s$^{-1}$ Mpc$^{-1}$, $\Omega_M$=0.3, $\Omega_{\Lambda}$=0.7, and k=0.

\section{Suzaku Observations and Data Reduction}\label{xray}
\subsection{Suzaku Observations}
We observed the three FeLoBALs in 2009 with the \suz\ X-ray observatory \citep{m07}, with exposure times ranging between 32--36~ks 
for each object. Table~\ref{suzstat} lists the observation log. During each observation, the standard data mode was used with the pointing 
observation mode. The three working X-ray Imaging Spectrometer (XIS) cameras (XIS0, XIS1, XIS3, Koyama et al.\ 2007) were operated in both the 3x3 
and 5x5 photon counting modes with a minimum time resolution of 8 seconds. \suz\ operates in a low-earth orbit, 550 km above the earth, 
completing one period in 96 minutes. The background in our observations is low and stable, without flares, consistent with most \suz\ observations \citep{b9}.

\begin{deluxetable}{lcccc}
\tablenum{1}
\tablewidth{0pt}
\tablecaption{The \suz\ Observation Log \label{suzstat}}
\tabletypesize{\scriptsize}
\tablehead{
\colhead{} & \colhead{Observation} & \colhead{Observation} & \colhead{3x3 mode} & \colhead{5x5 mode} \\
\colhead{Object} & \colhead{Start (UT)} & \colhead{Stop (UT)} & \colhead{exposure (s)} & \colhead{exposure (s)}
}
\startdata
J0943+5417 & 14:08:03, May 24, 2009 & 08:20:12, May 25 & 28,210.81 & 6,015.56 \\
J1352+4239 & 23:54:09, Jun 2, 2009 & 18:00:19, Jun 3 & 23,792.67 & 8,289.84 \\
J1723+5553 & 08:49:36, Jun 4, 2009 & 19:33:19, Jun 4 & 35,944.81 & \nodata \\
\enddata
\tablecomments{Only data for the 3x3 mode exists for J1723+5553.}
\end{deluxetable}

\subsection{Suzaku Data Reduction}
\label{sdr}
Starting with the cleaned event files, we visually inspect the data using \emph{DS9}\footnote{An astronomical imaging and data visualization application 
\citep{b8}} to check for any anomalies and ensure that our data extraction is completely contained within the \suz\ field. Using 
the standard aperture photometry, the half-power diameter region around each source is obtained using the \emph{Suzaku Technical 
Description}\footnote{Available at http://heasarc.gsfc.nasa.gov/docs/suzaku/prop\_tools/suzaku\_td/suzaku\_td.html.}, along with a 
corresponding background annulus. The background annuli are chosen individually for each mode (3x3, 5x5) to maximize the background area while 
keeping the inner radius large enough not to be contaminated by the source and the outer radius still contained within the field. The effective area 
for the background annulus is measured from the mean of the inner and outer radii of the background region. We then use the High Energy Astrophysics 
Science Archive Research Center (HEASARC) tool XSELECT in conjunction with these regions to filter the energy to the observed full (0.2--10 keV), soft 
(0.2--2 keV), and hard (2--10 keV) bands, extracting the counts and exposure times. We first 
combine the 3x3 and 5x5 mode data, then measure the source and background counts for the three CCDs, XIS0, XIS1, and XIS3, separately. 
We note that XIS1 is a back-illuminated instrument, which has a higher background and sensitivity at lower energy ranges. The count rates and 
uncertainties for XIS0 and 
XIS3, which are both front illuminated (FI) instruments, are found separately and then combined using the Bayesian estimation method to give a total FI count 
rate and error for the count rate. The final count rates and uncertainties are listed in Table~\ref{cntrate}.

We first examine the full band, and do not detect the sources; then we examine the soft band, with the same result. In the hard band, we do not 
detect the sources in most situations. We marginally detect SDSS~J1723+5553 ($2.6\sigma$ and $1.6\sigma$) in both of 
the FI CCDs in the observed 2--10 keV band. Combining the data from the two FI CCDs, we detect J1723+5553 at the 3$\sigma$ significant level 
(Table~\ref{cntrate}). The FI CCDs have comparable effective areas, and lower background rates at 2--10 keV than the BI CCD, which means they are more 
efficient at detecting faint sources in this energy band. Thus the non-detection of the source in the BI CCD (XIS1) is expected, given 
the low X-ray flux. Therefore, we conclude that the detection of J1723+5553 in the 2--10 keV band is real. We do not detect the other objects in 
any of the detectors.

Finally, we use the HEASARC tool PIMMS to convert the count rates into fluxes. Since the intrinsic column density of the quasars is unknown, we use 
a range of $10^{20}\le N_H \le 10^{25}$ \cmsq\ to calculate the unabsorbed, intrinsic fluxes. For the galactic column density, 
the \cite{b4} average $N_H$ values are used ($1.3\times10^{20}$ \cmsq, $1.21\times10^{20}$ \cmsq, and $3.08\times10^{20}$ \cmsq, for J09843+5417, 
J1352+4239, and J1723+5553, respectively). The photon index is taken to be $\Gamma$=1.9 (e.g., Reeves \& Turner 2000; Dai et al.\ 2004; 
Saez et al.\ 2008), equivalent to an energy index $\alpha$=0.9. The unabsorbed fluxes are calculated from the BI and FI rates 
separately using PIMMS and then the upper limits are combined using the Bayesian estimation method. We use just the FI rates in 
calculations for J1723+5553. Next, we estimate the monochromatic 2~keV (rest frame) luminosity using the unabsorbed fluxes (or flux limits) 
in the observed 0.2--2 and 2--10 keV bands. Using the power-law model, the flux, $F$, is given by, 
\begin{equation}
F = \int_{\nu_1}^{\nu_2} f_0 \nu^{-\alpha} d\nu, \label{eqnf}
\end{equation}
where $f_0$ is a constant found analytically for the 2--10 and 0.2--2 keV ranges.

From the simple relation in Equation \ref{eqnf}, we interpolate the value for the monochromatic flux density at the observed energy $E_X = 2~keV/(1+z)$ 
using,
\begin{equation}
\label{eqn2kev}
f_{\nu_X} = f_0 \nu_X^{-\alpha}.
\end{equation}
We find the 2 keV rest frame luminosity ($l_{2keV}$) from the corresponding observed fluxes using, 
\begin{equation}
\label{exrayl}
l_{\nu,rest} = \frac{f_{\nu,obs}}{(1+z)} 4 \pi D_L^2,
\end{equation}
where $\nu, obs = \nu / (1+z) $ and $D_L$ is the luminosity distance.  Since $f_{\nu,obs}$ is calculated for a range of intrinsic absorption, 
we obtain a range of monochromatic luminosities at 2~keV (Table~\ref{lum}).

\begin{deluxetable}{lccccc}
\tabletypesize{\scriptsize}
\tablenum{2}
\tablewidth{0pt}
\tablecaption{Count Rates and Uncertainties for Each Object\label{cntrate}}
\tablehead{
\multicolumn{1}{c}{} & \multicolumn{2}{c}{Observed 0.2--2 keV} & \multicolumn{1}{c}{} & \multicolumn{2}{c}{Observed 2--10 keV} \\ \cline{2-3} \cline{5-6}
\colhead{Object} & \colhead{FI count rate} & \colhead{BI count rate} & \colhead{} & \colhead{FI count rate} &\colhead{BI count rate}
}
\startdata
J0943+5417 & $<3.0\times10^{-4}$ & $<6.7\times10^{-4}$ & & $<3.3\times10^{-4}$ & $<8.5\times10^{-4}$ \\ 
J1352+4239 & $<2.1\times10^{-4}$ & $<9.0\times10^{-4}$ & & $<3.9\times10^{-4}$ & $<9.9\times10^{-4}$ \\
J1723+5553 & $<2.3\times10^{-4}$ & $<6.9\times10^{-4}$ & & $4.4\pm1.4\times10^{-4}$ & $<9.0\times10^{-4}$ \\
\enddata
\tablecomments{The 3$\sigma$ upper limits are reported here, and the 3$\sigma$ hardband detection of J1723+5553. The count rates are within the half-power 
diameter region of \suz, and are scaled to full power after the conversion to fluxes. The units are cnt~s$^{-1}$.}
\end{deluxetable}

\begin{deluxetable}{ccccccccccc}
\tabletypesize{\scriptsize}
\tablenum{3}
\tablewidth{0pt}
\tablecaption{2 keV rest frame Monochromatic flux and luminosity.\label{lum}}

\tablehead{
\multicolumn{2}{c}{} & \multicolumn{2}{c}{J0943+5417} & \colhead{} & \multicolumn{2}{c}{J1352+4239} & \colhead{} & \multicolumn{2}{c}{J1723+5553} \\ \cline{3-4} \cline{6-7} \cline{9-10}
\colhead{} & \colhead{Intrinsic} & \colhead{} & \colhead{} & \colhead{} & \colhead{} & \colhead{} & \colhead{} & \colhead{} & \colhead{} \\
\colhead{} & \colhead{$N_H$} & \colhead{$f_{X}$} & \colhead{$l_{X}$} & \colhead{} & \colhead{$f_{X}$} & \colhead{$l_{X}$} & \colhead{} & 
\colhead{$f_{X}$} & \colhead{$l_{X}$}
}
\startdata
& $10^{20}$ & $<1.4\times10^{-32}$ & $<4.8\times10^{26}$ & & $<1.0\times10^{-33}$ & $<2.7\times10^{26}$ & & $<9.3\times10^{-33}$ & $<2.7\times10^{26}$ \\
Extrapolated & $10^{21}$ & $<1.5\times10^{-32}$ & $<4.9\times10^{26}$ & & $<1.0\times10^{-32}$ & $<2.7\times10^{26}$ & & $<9.4\times10^{-33}$ & $<2.8\times10^{26}$ \\
from & $10^{22}$ & $<1.6\times10^{-32}$ & $<5.3\times10^{26}$ & & $<1.1\times10^{-32}$ & $<3.0\times10^{26}$ & & $<1.0\times10^{-32}$ & $<3.1\times10^{26}$ \\
0.2--2 keV & $10^{23}$ & $<2.9\times10^{-32}$ & $<9.7\times10^{26}$ & & $<2.2\times10^{-32}$ & $<6.0\times10^{26}$ & & $<2.0\times10^{-32}$ & $<6.0\times10^{26}$ \\
Observed & $10^{24}$ & $<5.6\times10^{-31}$ & $<1.9\times10^{28}$ & & $<6.2\times10^{-31}$ & $<1.7\times10^{28}$ & & $<4.9\times10^{-31}$ & $<1.4\times10^{28}$ \\
& $10^{25}$ & $<1.7\times10^{-23}$ & $<5.6\times10^{35}$ & & $<3.4\times10^{-22}$ & $<9.1\times10^{36}$ & & $<8.2\times10^{-23}$ & $<2.4\times10^{36}$ \\
\hline
& $10^{20}$ & $<1.8\times10^{-32}$ & $<5.9\times10^{26}$ & & $<1.9\times10^{-32}$ & $<5.0\times10^{26}$ & & $2.9\times10^{-32}$ & $8.5\times10^{26}$ \\
Extrapolated & $10^{21}$ & $<1.8\times10^{-32}$ & $<5.9\times10^{26}$ & & $<1.9\times10^{-32}$ & $<5.0\times10^{26}$ & & $2.9\times10^{-32}$ & $8.5\times10^{26}$ \\
from & $10^{22}$ & $<1.8\times10^{-32}$ & $<6.0\times10^{26}$ & & $<1.9\times10^{-32}$ & $<5.1\times10^{26}$ & & $2.9\times10^{-32}$ & $8.5\times10^{26}$ \\
2--10 keV & $10^{23}$ & $<1.9\times10^{-32}$ & $<6.4\times10^{26}$ & & $<2.0\times10^{-32}$ & $<5.5\times10^{26}$ & & $3.1\times10^{-32}$ & $9.2\times10^{26}$ \\
Observed & $10^{24}$ & $<3.5\times10^{-32}$ & $<1.2\times10^{27}$ & & $<3.9\times10^{-32}$ & $<1.1\times10^{27}$ & & $5.9\times10^{-32}$ & $1.7\times10^{27}$ \\
& $10^{25}$ & $<2.7\times10^{-31}$ & $<8.9\times10^{27}$ & & $<3.5\times10^{-31}$ & $<9.5\times10^{27}$ & & $5.0\times10^{-31}$ & $1.5\times10^{27}$ \\[-0.9ex]
\enddata
\tablecomments{\ Listed are the monochromatic fluxes and luminosities for the rest frame energy corresponding to 2 keV observed. The units are erg s$^{-1}$ 
cm$^{-2}$ Hz$^{-1}$ for flux, $f_X$, and \lumind\ for luminosity, $l_X$.  All fluxes and luminosities are upper limits, except for the 3$\sigma$ detection. }
\end{deluxetable}

\section{The 2500\AA\ Luminosity}\label{sdata}

\begin{figure}
\figurenum{1}
\begin{center}
\includegraphics[width=4.5 in]{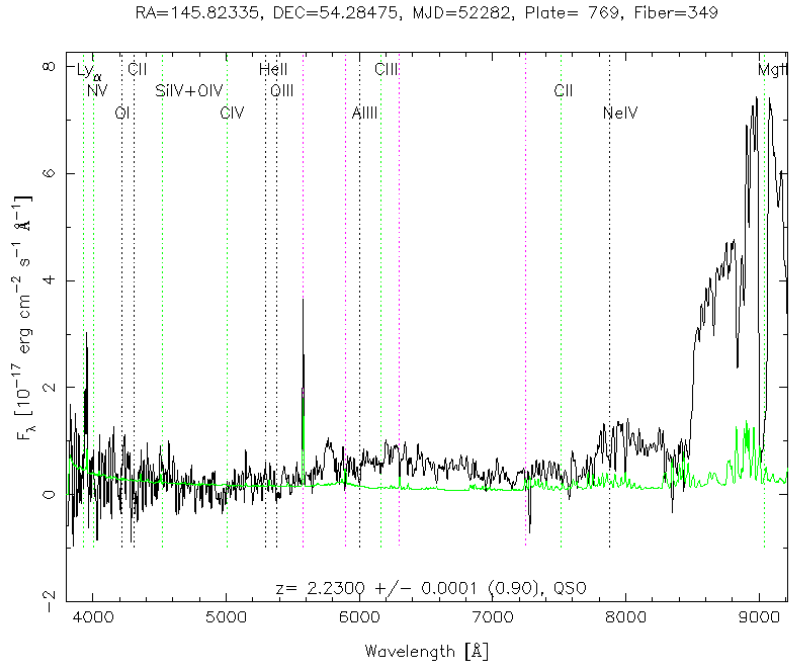}
\label{spec0943}
\caption{Observed optical spectrum of J0943+5417 from the \sdss.  The suppression of the spectrum here is a clear example of why 
it is difficult to correct the observed optical data.}
\end{center}
\end{figure}

We calculate the rest frame 2500\AA\ luminosity, $l_{2500}$, by extrapolating the observed \mass\ magnitudes (probing rest frame 4000--7000\AA\ for 
$z\sim2$ objects) using the mean quasar SED of \citet{b10}. Since all three objects are at $z\sim2$, the \mass\ bands sample the rest-frame optical bands of 
the FeLoBALs, where there is less spectral complexity than the rest-frame UV bands.  The \sdss\ (\emph{SDSS}) magnitudes sample the rest 
frame UV of the FeLoBALs, where it is extremely difficult to model the continuum due to the presence of complicated absorption troughs 
and dust extinction. Although the observed \mass\ wavelengths are further from the rest frame UV than the observed \emph{SDSS} 
wavelengths, we do not have to rely on a difficult to estimate model of the continuum. The observed optical spectra of J0943+5417 is 
almost completely suppressed (see Figure~\ref{spec0943}), which demonstrates why we use the \mass\ data that circumvents reliance on complex corrections.

We first correct for Galactic extinction caused by the Milky Way for the \mass\ magnitudes. This extinction is given for the \textit{u}-band in the 
\emph{SDSS} quasar catalog for each object (Schneider et al.\ 2007). From that value, we obtain the extinction in the \mass\ bands using the Milky Way 
extinction curve,
\begin{equation}
\frac{A_{\lambda}}{A_B}=\left(\frac{E(\lambda-V)}{E(B-V)} + R_V\right)\frac{1}{1+R_V},
\label{extinct}
\end{equation}
where the values of color excesses, $E(\lambda-V)$, and $R_V=3.08$ are taken from \citet{b3}. After correcting for Galactic extinction, we calculate 
the observed flux densities at the central $J$, $H$, and $K_s$ wavelengths, and obtain the rest frame monochromatic luminosities at optical bands 
(at $\sim$6,900\AA) using Equation~\ref{exrayl}. We correct for the intrinsic dust extinction in these quasars assuming $E(B-V)=0.077$ 
with an SMC-like extinction curve, found by \citet{b11} to be a good approximation for LoBALs. Since FeLoBALs have higher dust extinctions 
than LoBALs, we possibly under-correct the intrinsic dust extinction for our targets. Finally, we extrapolate the rest-frame optical 
luminosities to $l_{2500}$ using the mean quasar SED of \citet{b10}. The parameters and intermediate results used in all these calculations 
are listed in Table \ref{params}.  The results obtained from $J$, $H$, and $K_s$ bands are consistent within 20\% for J0943+5417 and J1352+4239, 
suggesting no significant complexity in the corresponding spectra. For J1723+5553, the 2500\AA\ luminosities extrapolated from $J$ and $H$ bands 
are consistent within 20\%; however, the value extrapolated from the $K_s$ band is twice as large. It is possible that the observed $K_s$ 
band is contaminated by a strong $H_{\alpha}$ emission line. For all three objects, we use the average of the $l_{2500}$ values extrapolated 
from the $J$, $H$, and $K$ bands, in our following calculations.

\begin{deluxetable}{lccccccccc}
\tabletypesize{\scriptsize}
\rotate
\tablenum{4}
\tablewidth{0pt}
\tablecaption{Rest Frame 2500\AA\ Luminosity\label{params}}
\tablehead{
\colhead{Object} & \colhead{z} & \colhead{$A_u$} & \colhead{$K_s$} & \colhead{$f_{21,590\text{\AA}}$}  
& \colhead{$l_{\sim6900\text{\AA}}$} & \colhead{$l_{2500}$ from $J$} & \colhead{$l_{2500}$ from $H$} & \colhead{$l_{2500}$ from $K_s$} \\
\colhead{} & \colhead{} & \colhead{} & \colhead{magnitude} & \colhead{\flux Hz$^{-1}$} & \colhead{\lumin Hz$^{-1}$} 
& \colhead{\lumin Hz$^{-1}$} & \colhead{\lumin Hz$^{-1}$} & \colhead{\lumin Hz$^{-1}$} \\
}
\startdata
J0943+5417 & 2.23 & 0.063 & 14.251 & 1.408$\times10^{17}$ &  1.6$\times10^{32}$ & 9.1$\times10^{31}$ & 1.3$\times10^{32}$ & 1.6$\times10^{32}$ \\
J1352+4239 & 2.04 & 0.055 & 13.861 & 2.002$\times10^{17}$ &  2.0$\times10^{32}$ & 1.5$\times10^{32}$ & 1.5$\times10^{32}$ & 1.8$\times10^{32}$ \\
J1723+5553 & 2.11 & 0.178 & 14.097 & 1.804$\times10^{17}$ &  1.9$\times10^{32}$ & 7.8$\times10^{31}$ & 9.5$\times10^{31}$ & 1.6$\times10^{32}$ \\
\enddata
\tablecomments{Statistics for all three objects. The redshift, $z$, and galactic extinction, $A_u$, are taken from \sdss; $K_s$ is the 
observed magnitude from \mass; $f_{21,590}$ is the observed flux in the $K_s$ band; $l_{\sim6900\text{\AA}}$ is the luminosity in the rest 
frame wavelengths of the $K_s$ band for each object; The last three columns contain the luminosity at 2,500\AA\ as extrapolated from 
the $J$, $H$, and $K_s$ bands.} 
\end{deluxetable}

\section{Results}
After calculating the intrinsic monochromatic luminosities at 2500\AA\ and 2~keV rest frame, we measure the broadband spectral index between the 
UV and X-ray bands, \aox $= 0.3838\times\log{(l_{2 keV}/l_{2500})}$, and then compare our \aox\ values to those of the large sample of 333 
normal quasars from \citet{b6}. The sample was chosen to cover as much of the $l-z$ plane as possible.  Other studies including FeLoBALs 
\citep{br05,b21} found a slightly different trend.  \citet{br05} assumed a galactic column density using \citet{s98} and found lower than expected 
intrinsic \aox\ values for their two FeLoBALs.  They lie right on the bottom edge of and just below ($\Delta$\aox=0.3) the scatter of expected \aox\ for 
the $l_{2500}\sim$\aox\ relation from \citet{b6}. \citet{b21} compared the observed \aox\ of their FeLoBALs with the \citet{b6} study, and found  
the observed \aox\ to be $\sim0.48$ below the expected scatter. From this they inferred an intrinsic column density. We follow the convention 
of \citet{b6} and use 2500\AA\ and 2~keV, although \aox\ can be defined using other optical wavelengths and X-ray energies, 
as shown by \citet{y10}. In this section, we first examine the dependency of \aox\ on the intrinsic absorption in our FeLoBALs, using \aox\ 
calculated independently from the soft X-ray (0.2--2 keV, observed) and the hard X-ray (2--10 keV, observed) spectra. Then we analyze \aox\ as 
a function of UV and X-ray luminosities, and draw conclusions about the intrinsic column density.

\subsection{Intrinsic Absorption and $\alpha_{ox}$}
\begin{figure}[ht]
\figurenum{2}
\begin{center}
\subfigure[J0943+5417]{\includegraphics[angle=90,width=3.0in,clip,trim=0in 0in 0in -0.3in]{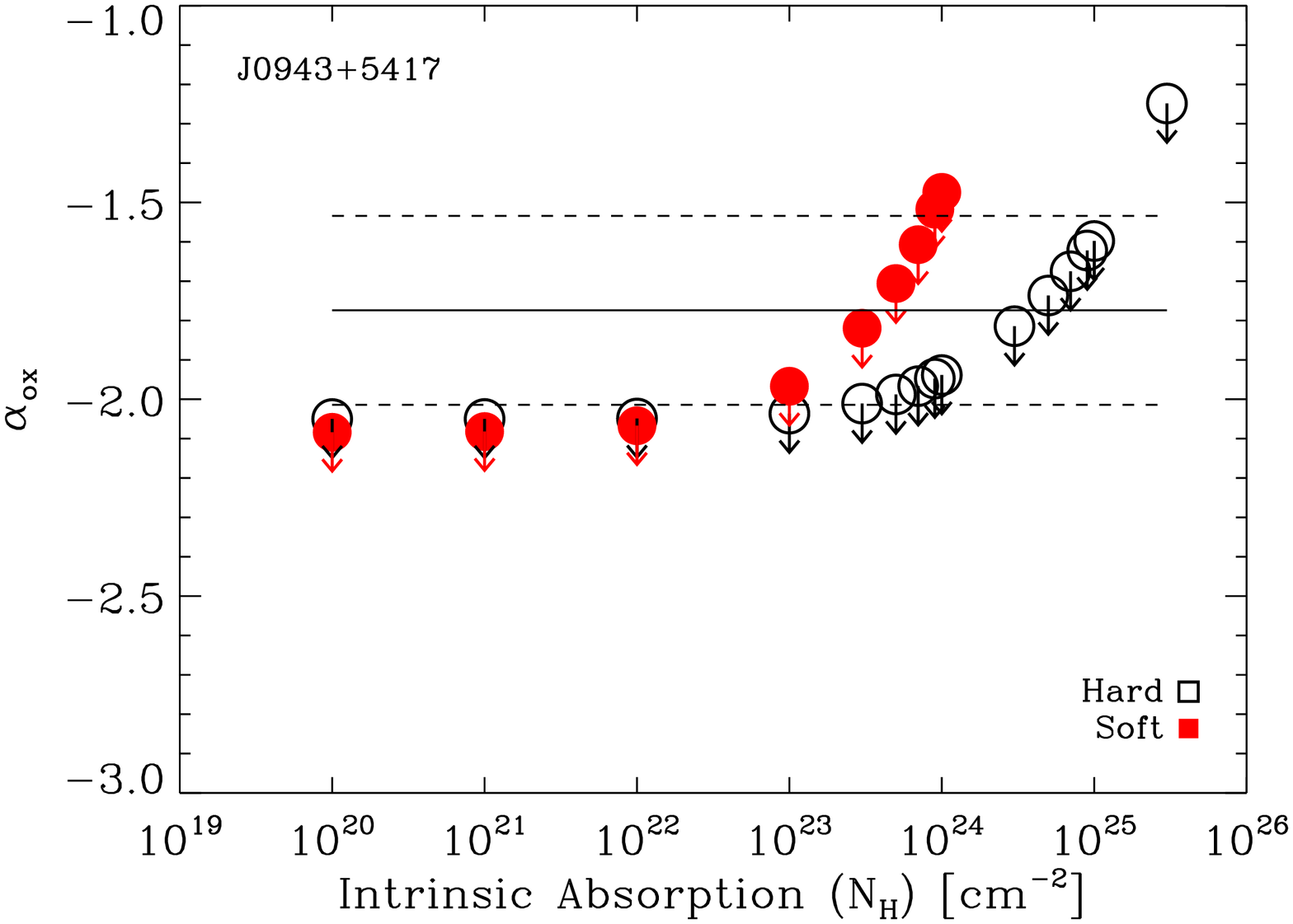}\label{fig:subfig1}}
\subfigure[J1352+4239]{\includegraphics[angle=90,width=3.0in,clip,trim=0in 0in 0in -0.3in]{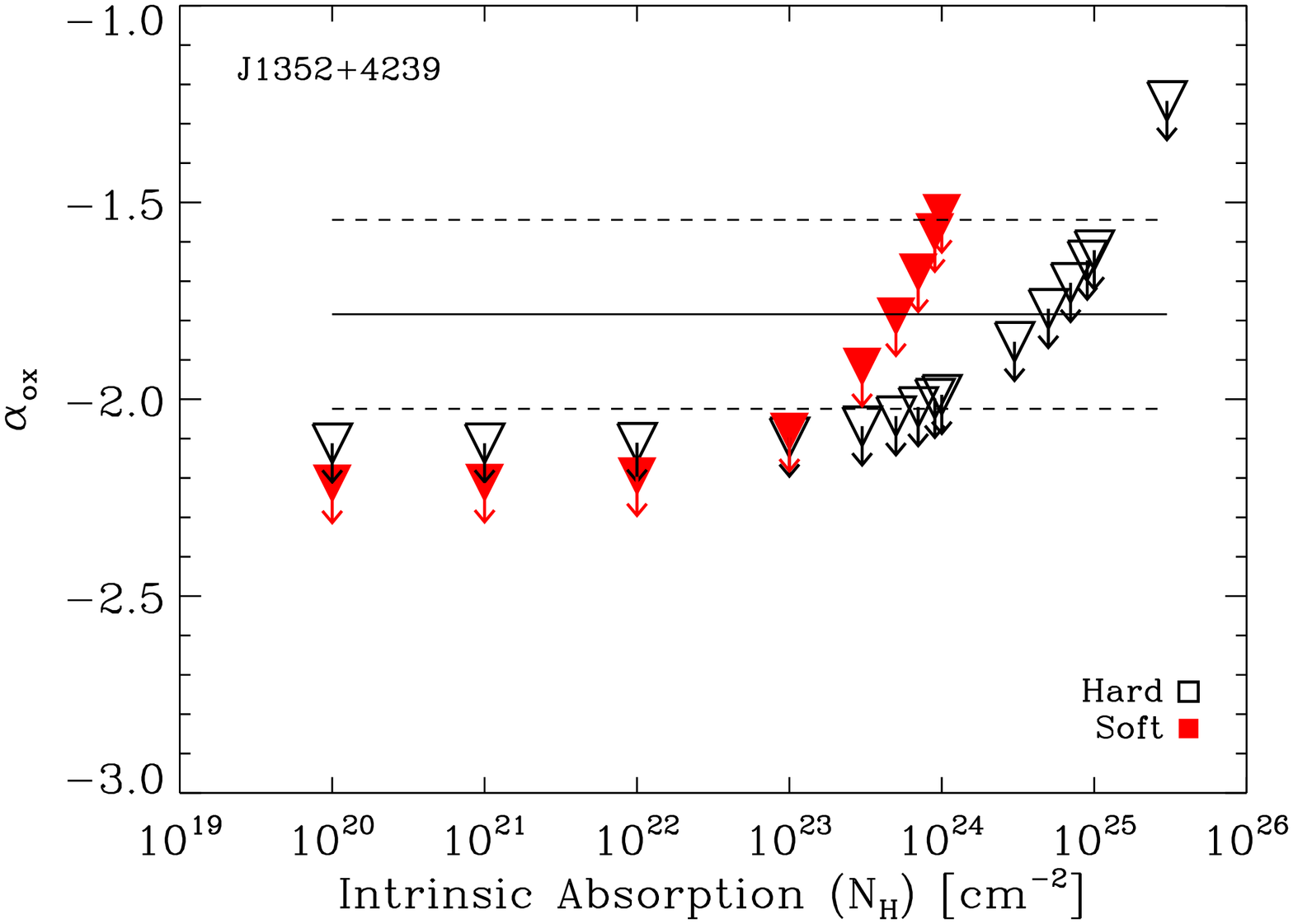}\label{fig:subfig2}}
\subfigure[J1723+5553]{\includegraphics[angle=90,width=3.0in,clip,trim=0in 0in 0in -0.3in]{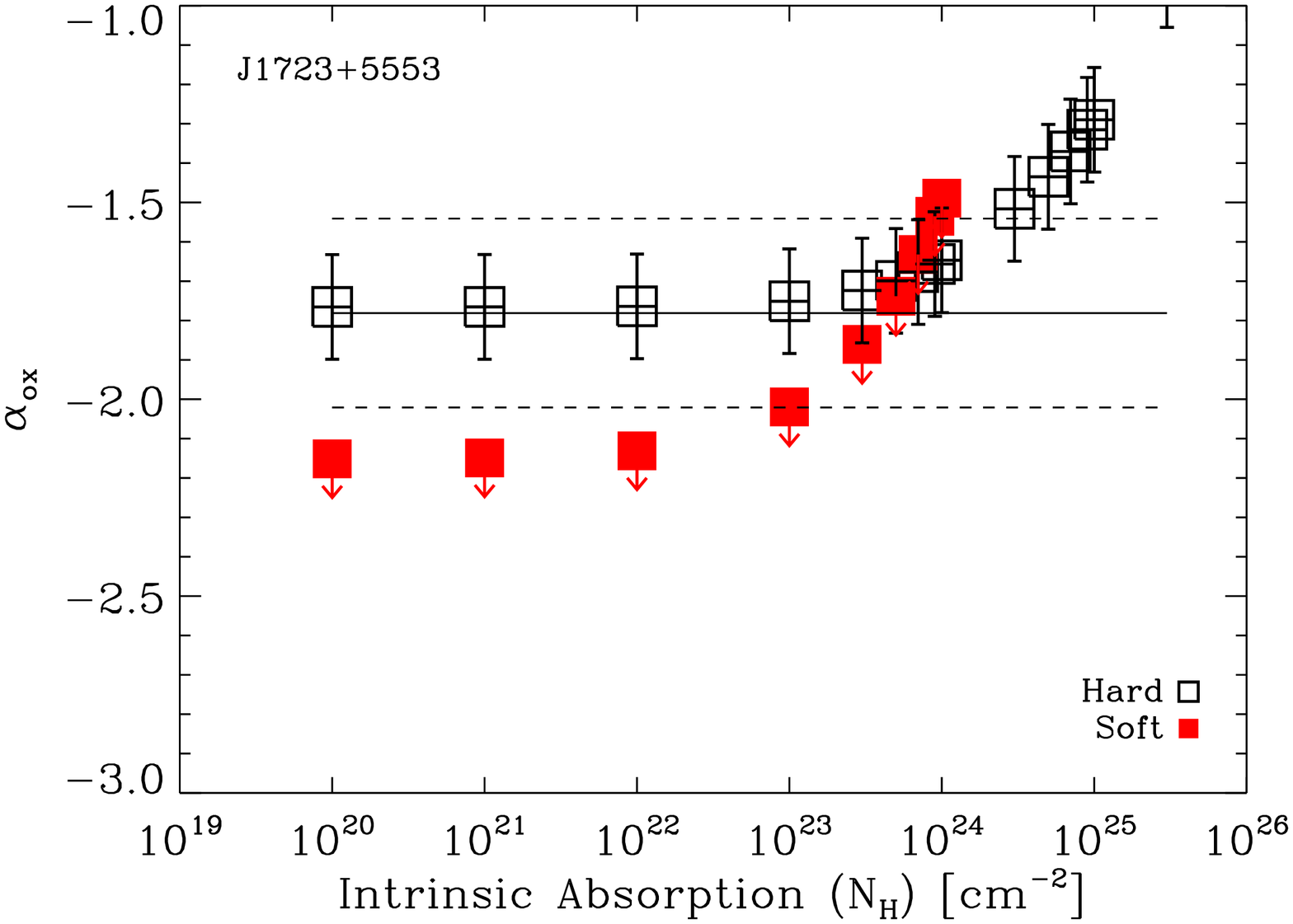}\label{fig:subfig3}}
\caption{Dependence of \aox\ on Intrinsic Absorption ($N_H$).  Values calculated from the 2--10 keV range are shown in black, while values calculated 
from the 0.2--2 keV range are shown in red filled symbols.  The expected \aox\ from $l_{2500}$ is depicted by the solid line, with the expected scatter 
depicted by the dashed lines.}
\end{center}
\label{intnh}
\end{figure}

We first study the relation between \aox\ and the intrinsic absorption in Figure \ref{intnh}. Each sub-figure corresponds to one FeLoBAL, 
and shows the entire input range for intrinsic column density. The values calculated from the 2--10 keV (hard) spectrum are depicted in black, 
and the 0.2-2 keV (soft) spectrum are depicted by red, filled symbols. Where monochromatic luminosities at 2 keV are upper limits due to the 
non-detection of a source, \aox\ is also only an upper limit, depicted in all Figures by arrows. Included are solid lines for 
the expected \aox$\sim l_{2500}$ relation, as found by \citet{b6}, and dashed lines to depict the expected scatter.

At 2 keV the intrinsic luminosity and its uncertainty are dependent on the column density within the FeLoBAL itself, which causes intrinsic 
absorption that must be corrected for. We use a range $10^{20} \le N_H \le 10^{25}$ \cmsq\ for intrinsic column densities since the 
true quantity is unknown. This gives a different value of intrinsic luminosity, and therefore \aox, for each column density. The 
luminosities at 2500\AA\ are extrapolated from the rest frame optical luminosities, using a normal quasar SED from \citet{b10}, where we have 
corrected both the intrinsic extinction and Galactic extinction due to dust as discussed in \S~\ref{sdata}. We find the uncertainty in the 
luminosities to be $\sim$20\%.

In the two non-detected FeLoBALs (J0943+5417 and J1352+4239), for column densities up to $1\times10^{23}$ \cmsq, the soft limit is more 
stringent for \aox, although the difference between the two 
limits diverges by $\le0.2$. At higher column densities, the hard spectrum is the appropriate limit to use. The divergence 
between the soft and hard limits is increasingly larger with increasing column density. J1723+5553 is a different story: only 
at $N_H\ge 6\times10^{23}$ \cmsq\ are the upper limits from the soft X-ray spectrum 
consistent with the 3$\sigma$ detection in the hard X-ray spectrum. The 3$\sigma$ detection and its uncertainties for \aox\ calculated from 
the hard X-rays are $\ge$0.3 above the upper limits given by the soft X-ray energy range for column densities $< 6\times$10$^{23}$ \cmsq. Thus, 
the X-ray hardness ratio of J1723+5553 constrains the intrinsic column density to $N_H\ge 6\times10^{23}$ \cmsq.

\subsection{Monochromatic Luminosities vs. \aox}\label{mla}
\begin{figure}[ht]
\figurenum{3}
\begin{center}
\includegraphics[angle=90,width=4.5in,clip,trim=0in 0in 0in -0.3in]{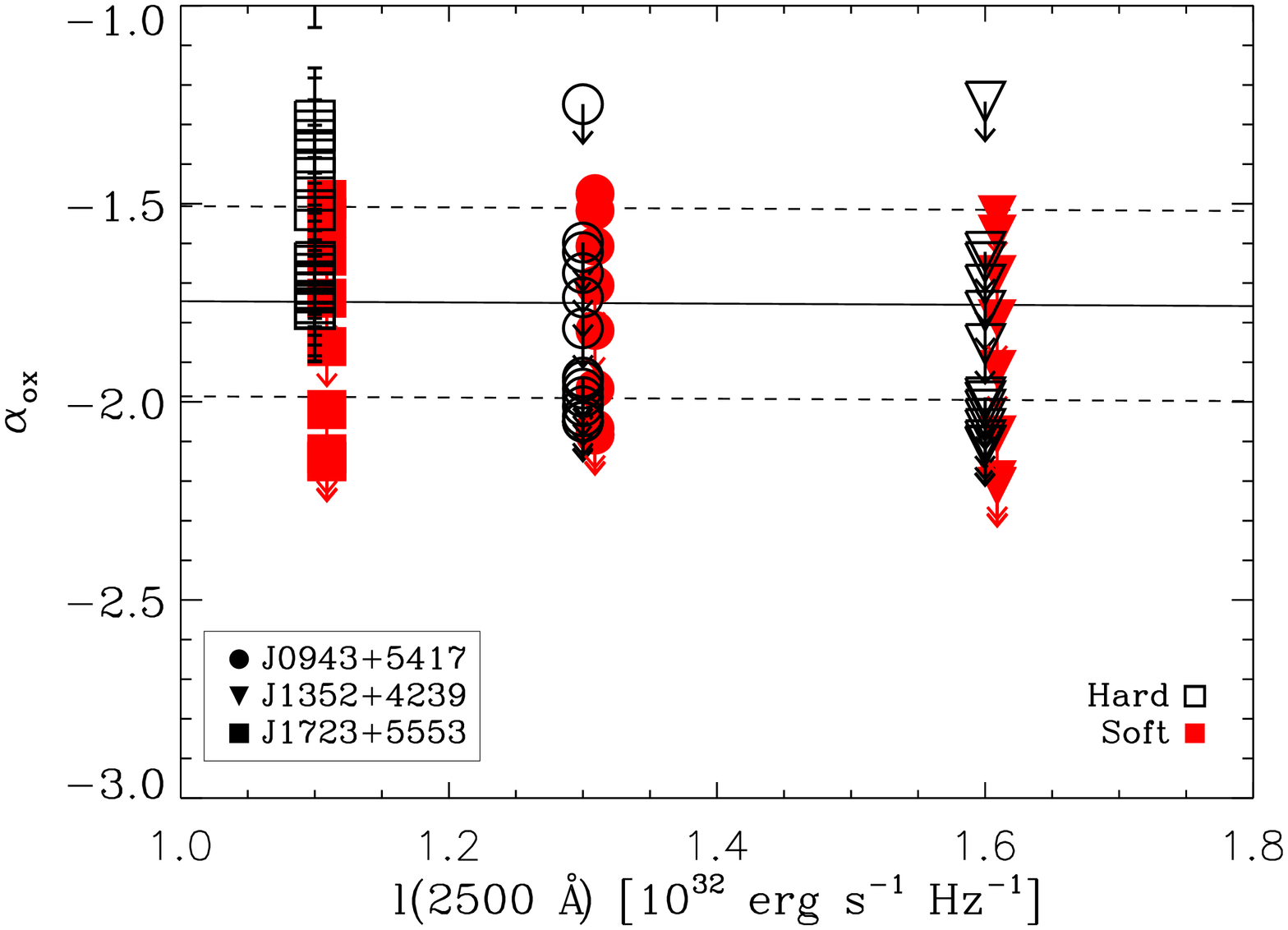}
\caption{$\alpha_{ox}$ vs. $l_{2500\text{\AA}}$ monochromatic luminosity. Values calculated from the 2--10 keV range are shown in black, while values calculated 
from the 0.2--2 keV range are shown in red filled symbols and shifted to the right for clarity. The solid and dashed lines represent 
the mean and scatter (0.24) of the relation from \citet{b6}.}
\end{center}
\label{l2500}
\end{figure}

Figure \ref{l2500} shows the correlation between $l_{2500}$ and \aox, with the best-fit linear regression for normal AGN found by \citet{b6} 
plotted. This relation is given by \aox$=(-0.137\pm0.008)$log$(l_{2500})+(2.638\pm0.240)$ \citep{b6}. The solid line shows the 
expected \aox, and the scatter is depicted by the dashed lines. Column density increases as \aox\ becomes less negative. The upper 
limits on \aox\ from the soft X-ray spectrum are shifted slightly to the right of the hard X-ray upper limits in Figure~\ref{l2500} 
for ease of viewing. The upper limits plotted for J0943+5417 and J1352+4239 
are consistent with the normal AGN in the scatter of $\pm$0.24 for $N_H \ge 4\times10^{23}$ and $N_H \ge 8\times10^{23}$ \cmsq, 
respectively. This range falls in the column density region ($N_H \ge 1\times10^{23}$) where the hard spectrum provides the appropriate 
upper limit. For J1723+5553, combining the constraint from X-ray hardness ratio ($\nh > 6\times10^{23}\cmsq$), the values of \aox\ 
fall within the expected scatter if the intrinsic absorption is in the range of $6\times10^{23} \le N_H \le 3\times10^{24}$ \cmsq.

\begin{figure}[ht]
\figurenum{4}
\begin{center}
\subfigure[]{\includegraphics[angle=90,width=3.0in,clip,trim=0in 0in 0in -0.3in]{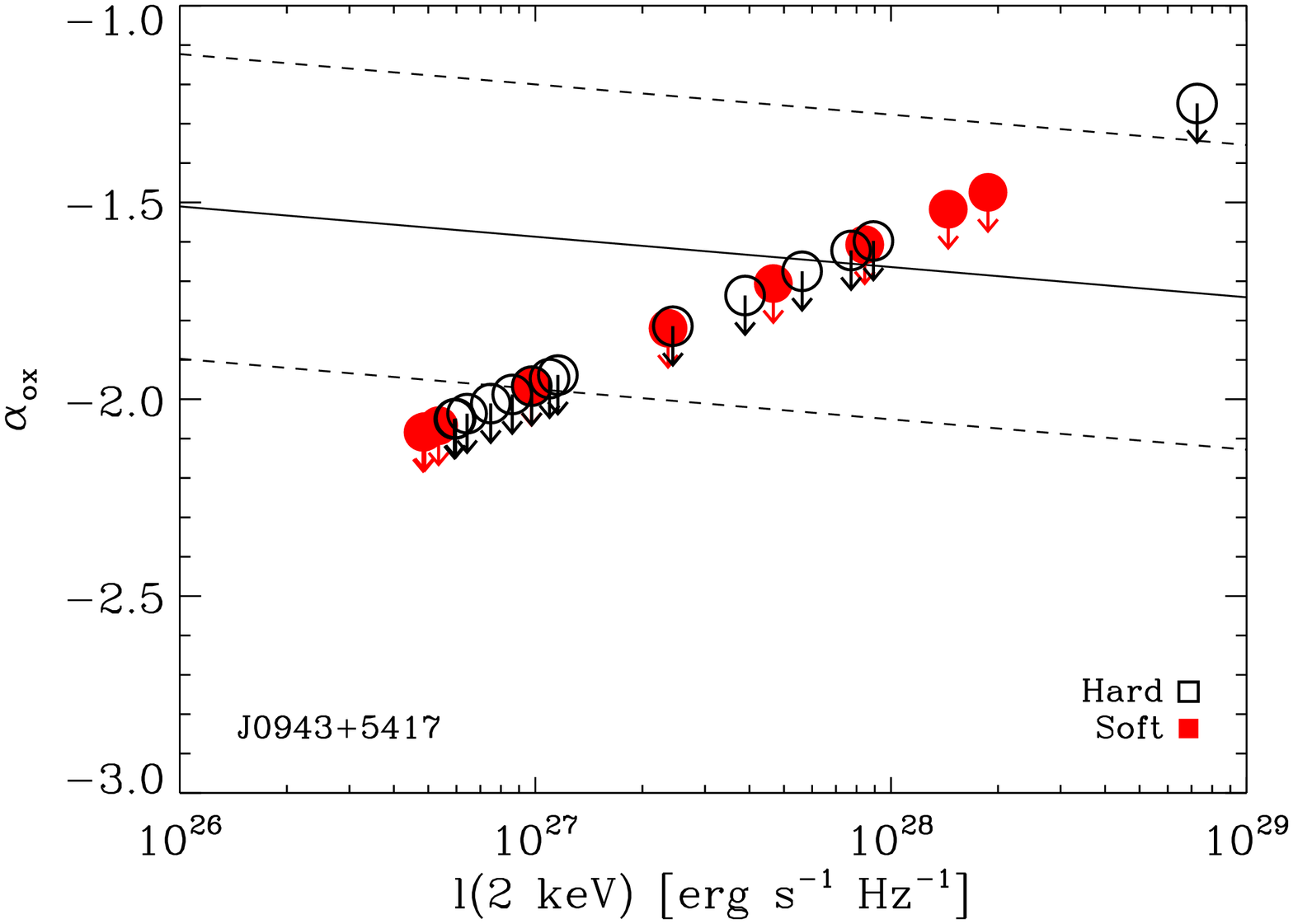}}
\subfigure[]{\includegraphics[angle=90,width=3.0in,clip,trim=0in 0in 0in -0.3in]{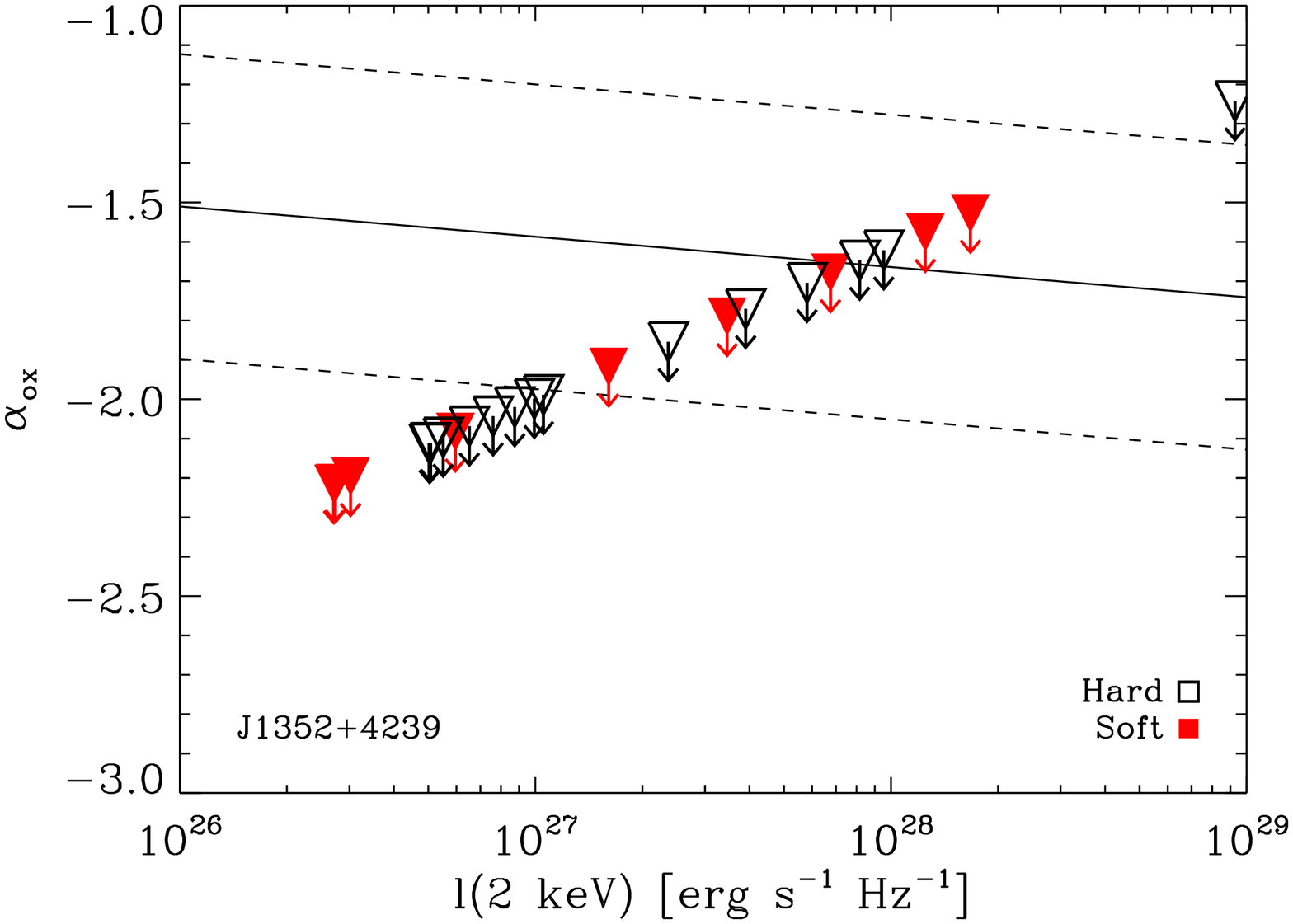}}
\subfigure[]{\includegraphics[angle=90,width=3.0in,clip,trim=0in 0in 0in -0.3in]{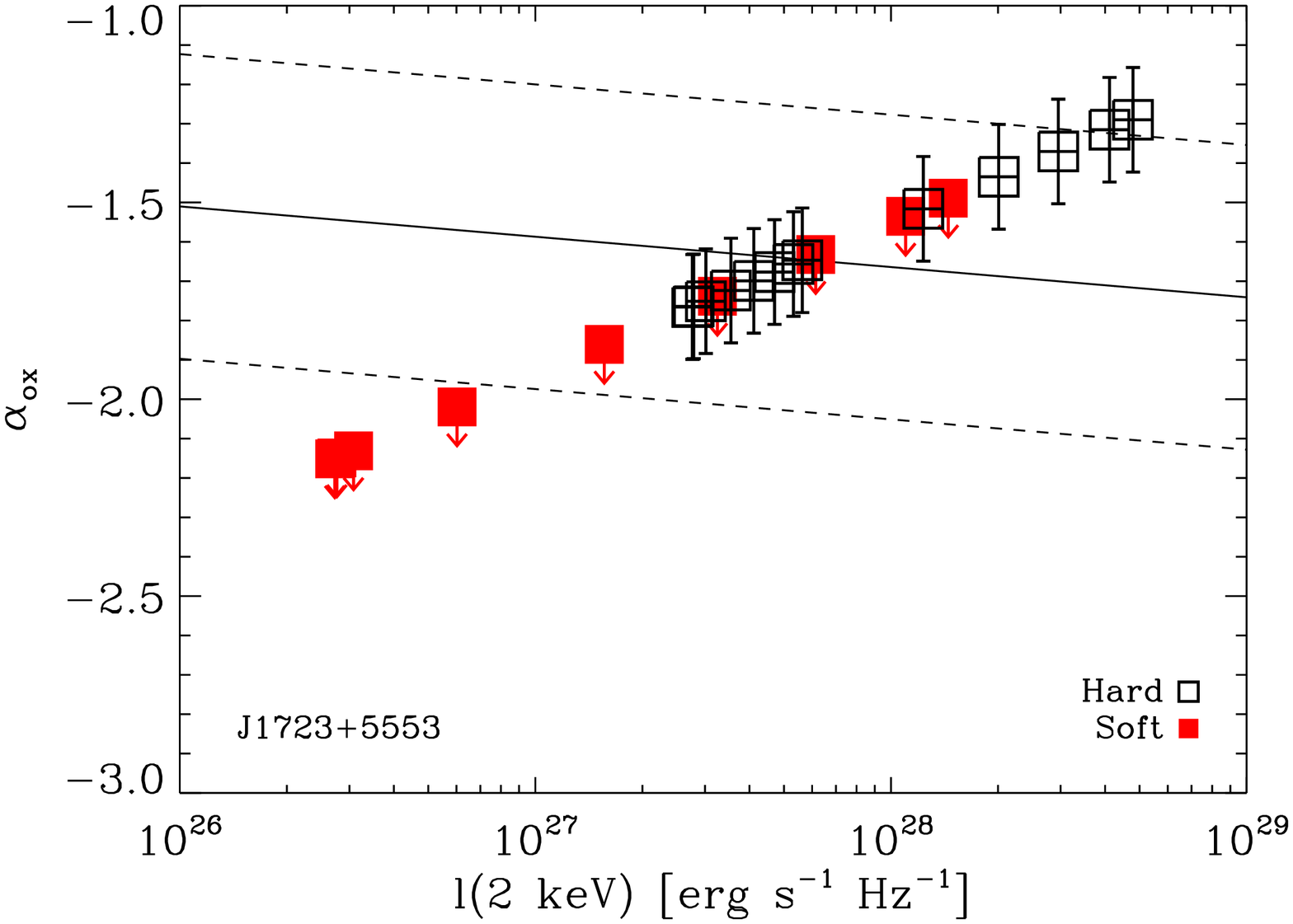}}
\caption{$\alpha_{ox}$ vs. $l_{2 keV}$ monochromatic luminosity. Values calculated from the 2--10 keV range are shown in black, while values calculated
from the 0.2-2 keV range are shown in red filled symbols. The solid and dashed lines represent  the mean and scatter (0.387) of the relation as seen in \citet{b6}.}
\end{center}
\label{l2kev}
\end{figure}

Figure~\ref{l2kev} shows the correlation between $l_{2keV}$ and \aox, again with the best-fit linear regression and expected scatter, using 
the relation, \aox$=(-0.077\pm0.015)$log$(l_{2 keV})+(0.492\pm0.387)$, from \citet{b6}. Column density increases from left to right, 
corresponding to rising intrinsic luminosity. For column densities $N_H \ge 7\times10^{23}$ \cmsq\ for J0943+5417 and $N_H \ge 2\times10^{24}$ 
\cmsq\ for J1352+4239, \aox\ is consistent with the normal AGN sample. The upper limits given by the soft spectrum fall within (and above) the 
scatter in the high column density region, where the hard spectrum gives the better constraint. Our 3$\sigma$ detection of J1723+5553 
indicates that for column densities $N_H \le 8\times10^{24}$ \cmsq, J1723+5553 is consistent with the SEDs of normal AGNs. 

Therefore, combining all constraints, J1723+5553 is consistent with normal AGN for column densities $6\times10^{23}\le N_H\le 3\times10^{24}$ \cmsq, 
while \aox\ for J0943+5417 and J1352+4239 are consistent with normal AGN for $N_H \ge 7\times10^{23}$ \cmsq\ and $N_H \ge 2\times10^{24}$ \cmsq, respectively.  

\section{Discussion}
\subsection{Column Densities and Intrinsic X-ray Luminosities}\label{scd}
We find significantly high column densities for all three FeLoBALs studied, $6\times10^{23} \le N_H \le 3\times10^{24}$ \cmsq\ for J1723+5553, 
$\nh \ge 7\times10^{23}$ \cmsq\ for J0943+5417, and $\nh \ge 2\times10^{24}$ \cmsq\ for J1352+4239. Only a small sample of FeLoBALs has been 
studied in X-rays previously. We summarize all the measurements in Table~\ref{density}, which is divided 
into two subsamples: those that are radio loud, and those that are not. The radio loud objects tend to have lower column densities, which 
could be due to geometry or evolution. The X-ray emission may be linked to the jets as well as from the corona right above the central portion of the 
accretion disk \citep[e.g.,][]{m09}. The BAL wind (see \citet{ge07} for a detailed illustration), would 
then only absorb X-ray emission from the accretion disk region. We 
do not know for certain what complicated structure dependence for X-ray emission may exist in a radio loud FeLoBAL, but the column 
density results are still important and thus we include them for completeness. Our three FeLoBALs are radio quiet, and our constraints 
are consistent with the subsample that is not known to be radio loud. These objects in general are found 
to have \nh\ on the order of a few $10^{23}$ \cmsq\ or higher. Notably, the most recent study by \citet{b21} found column densities for 
J2215-0045 and J0300+0048 to be $\nh \sim10^{24}$, with J2215-0045 just above the range we calculate for J1723+5553, but still consistent with our 
\nh\ for the two non-detections. The most recent calculation of column density for MrK 231 by \citet{br04}, using \emph{BeppSAX}, found 
\nh$\sim2\times10^{24}$ \cmsq, in agreement with all three of our objects. \citet{b16} cite lower column densities ($\nh \ge 6.5\times10^{22}$ \cmsq), 
but they also noted that this is an average expected column density for a HiBAL, and that \nh\ for a LoBAL would be higher.

The fact that our 3$\sigma$ detection in J1723+5553 is in the observed 2--10 keV (rest frame 6--30 keV) hard band rather than the soft X-ray 
band is significant, as this is expected from highly absorbed X-ray emission from the corona above the SMBH. The non-detection of the source in the 
observed 0.2--2 keV (rest frame 0.6--6 keV) band is consistent with our interpretation that we detect the direct X-ray emission in the 
observed 2--10 keV band, as high column densities will suppress the soft X-rays more than the hard X-rays. This is further supported 
by the fact that none of these objects is detected, even at the 3$\sigma$ level of significance, for the broadband observed 
0.2--10 keV (rest frame 0.6--30 keV) energy range. If the observed 2--10 keV spectrum is reflected or scattered into our line of sight, the 
X-ray spectrum will be a flat power-law, and we would also detect the source in the observed 0.2--2 keV band, which has a larger 
collecting area for \suz. Combining the measurements from this paper and those from \citet{br04, b21}, it is most likely that the 
intrinsic \nh\ column density absorbing the X-ray emission in a radio-quiet FeLoBAL is in the range from $\sim 10^{24}$ up to $10^{25}$ 
\cmsq. This is about 1--2 orders of magnitude higher than the X-ray absorption range in BALQSOs ($N_H\sim10^{22}-10^{24}$ \cmsq, 
e.g., Gallagher et al.\ 2006).

Using \aox\ to provide limits on intrinsic column density also allows us to provide limits on the intrinsic X-ray luminosity. Using the 
combined constraints summarized at the end of Section~\ref{mla}, the intrinsic X-ray luminosity for J0943+5417 is $l_{2{\rm \, keV}}\ge 9.8\times10^{26}$ \lumind; 
for J1352+4239 it is $l_{2{\rm \, keV}}\ge 1.7\times10^{27}$ \lumind; for J1723+5553 it is $4.4\times10^{27}\le 
l_{2{\rm \, keV}}\le 1.2\times10^{28}$ \lumind. Compared 
to the normal sample of AGN from \citet{b6}, these luminosities fall into the middle to high range of X-ray luminosities for normal quasars. This 
is predicated on the fact that we assumed an intrinsic SED for a normal AGN. The broadband luminosities for these objects in the soft band are 
$L_{0.6-6{\rm \, keV}} \ge 3.8\times10^{44}$ \lumin, $L_{0.6-6{\rm \, keV}} \ge 6.9\times10^{44}$ \lumin, and $1.8\times10^{45} 
\le L_{0.6-6{\rm \, keV}} \le 4.9\times10^{45}$ \lumin, for J0943+5417, J1352+4239, and J1723+5553, respectively. The corresponding hard band 
luminosities are $L_{6-30{\rm \, keV}} \ge 3.2\times10^{44}$ \lumin, $L_{6-30{\rm \, keV}} 
\ge 5.85\times10^{44}$ \lumin, and $1.5\times10^{45} \le L_{6-30{\rm \, keV}} \le 4.2\times10^{45}$ \lumin.

\begin{deluxetable}{lccccc}
\tabletypesize{\scriptsize}
\tablenum{5}
\tablewidth{0pt}
\tablecaption{X-ray Observations of FeLoBALs \label{density}}
\tablehead{\multicolumn{5}{c}{FeLoBALs} \\
\hline \\[-2ex]
\colhead{Object} & \colhead{Detection} & \colhead{Observatory} & \colhead{$N_H$ (\cmsq)} & \colhead{Reference}}
\startdata
J0943+5417\tablenotemark{a} & no & \suz & $\ge 7\times10^{23}$ & This paper \\
J1352+4239 & no & \suz & $\ge 2\times10^{24}$ & This paper \\
J1723+5553 & no & \suz & $6\times10^{23}-3\times10^{24}$ & This paper \\
SDSS J0300+0048 & no & \chandra & $\ge1.8\times10^{24}$ & \citet{b21} \\
SDSS J2215-0045 & no & \chandra & $\ge3.4\times10^{24}$ & \citet{b21} \\
Mrk 231 & yes & \emph{BeppoSAX} & $\sim2\times10^{24}$ & \citet{br04} \\
Q0059-2735 & no & \chandra & $\ge6.5\times10^{22}$\tablenotemark{b} & \citet{b16} \\
\hline \\[-2.5ex]
\hline \\[-2ex]
\multicolumn{5}{c}{Radio Loud FeLoBALs} \\
\hline \\[-2ex]
\colhead{Object} & \colhead{Detection} & \colhead{Observatory} & \colhead{$N_H$ (\cmsq)} & \colhead{Reference}\\
\hline \\[-2ex]
SDSS J1556+3517 & yes & \chandra & $< 9.6\times10^{23}$ & \citet{kb09} \\
SDSS J2107-0620 & yes & \chandra & $< 9.6\times10^{23}$ & \citet{kb09} \\
SDSS J1044+3656 & yes & \chandra & $< 9.6\times10^{23}$ & \citet{kb09} \\
SDSS J0814+3647 & yes & \chandra & \nodata & \citet{m09} \\
SDSS J2107-0620 & no & \emph{XMM-Newton} & $4\times10^{23}$ & \citet{w08} \\
FIRST 1044+3517 & yes & \chandra & $\sim3\times10^{23}$ & \citet{br05} \\
FIRST 1556+4517 & yes & \chandra & $\sim3\times10^{23}$ & \citet{br05} \\
FTM 1004+1229 & yes & \chandra & $2.8\times10^{23}$ & \citet{u05} \\
FTM 1036+2828\tablenotemark{c} & yes & \chandra & $3.8\times10^{22}$ & \citet{u05} \\
FTM 0830+3759 & yes & \chandra & $2.7\times10^{22}$ & \citet{u05} \\
\enddata
\tablecomments{Previous studies of FeLoBALs in the X-ray. We have added column density constraints when the study provided them.}
\tablenotetext{a}{Radio moderate.}
\tablenotetext{b}{\citet{b16} note that this should represent an average HiBAL, and would be higher for a LoBAL.}
\tablenotetext{c}{Sometimes classified as a mini-BAL.}
\end{deluxetable}

\subsection{Thomson Scattering Implications}
\label{tci}
Such high column densities are interesting physically, as they are either in or on the cusp of a regime where the Thomson scattering cross-section,
$\sigma_T = 6.65\times10^{-25}$ cm$^2$, will be large enough to scatter incident photons. Although we probe the rest frame quasar spectra in 
the rest frame 6--30 keV (observed 2--10 keV) band, the Klein-Nishina correction to the Thomson scattering cross section is still 
negligible. For J1723+5553, using the expected \aox\ from the \aox--$l_{2kev}$ relation, the
optical depth is $\tau \sim 1$. For J0934+5417 and J1352+4239, the expected \aox\ value from the \aox--$l_{2keV}$ relation gives an 
intrinsic $\nh \sim 8.5\times10^{24}$ \cmsq, which indicates a Thomson scattering optical depth of $\tau \sim 6$. Such a large optical 
depth will also significantly modulate the UV flux ($\sim$400 times dimmer); however, these objects are already more luminous in the 
UV than most of the AGN in the \citet{b6} study. As another check, we calculate the black hole mass of the quasars assuming they are emitting 
at 1/3 of the Eddington luminosity (e.g., Shankar et al.\ 2010). We extrapolate our rest frame luminosities to 5100\AA\ and assume a 
bolometric correction of 10.33 for 5100\AA\ \citep{b10} before using Equation~\ref{mbh} to find the black hole masses.
\begin{equation}
\label{mbh}
M_{BH}\simeq\frac{1}{3}\left(\frac{L_{Edd}}{1.5\times10^{38}}M_{\odot}\right)\simeq \frac{3L_{bol}}{1.5\times10^{38}}M_{\odot}
\end{equation}
For the values reported here, all three FeLoBALs have $M_{BH}\sim1.4\times10^8M_{\odot}$. 
Even at the lower end of the column density constraints, $\nh \ge 6\times10^{23}$ \cmsq, the optical depth would be $\tau \sim .4$, 
which would dim the intrinsic luminosity by one and a half times. At the column density corresponding to the expected \aox\ for J0943+5417 and 
J1352+4239, $\sim 8.5\times10^{24}$ \cmsq\ (from the \aox--$l_{2kev}$ relation, and still within the limits for J1723+5553), the optical depth 
of $\tau \sim 6$ would mean the intrinsic luminosity would be increased by 400 times, giving a black hole mass of $M_{BH}\sim3.2\times10^{12}M_{\odot}$. 
Such a large mass is not a physical possibility; therefore, it is unlikely that these objects are 
super-luminous in the UV, and we only see the tiny fraction of Thomson scattered emission.

Since the X-ray emission region is expected to be smaller than the UV emission region from either variability (e.g., Chartas et al.\ 2001) 
or quasar microlensing (e.g., Dai et al.\ 2003, 2010a; Pooley et al.\ 2007; Morgan et al.\ 2008; Chartas et al.\ 2009) arguments, it is instead 
possible that the X-ray absorbing material is located between the X-ray and UV emission region and only covers the X-ray emitting region. This 
is consistent with the disk wind models of \citet{ge07} and Murray et al.\ (1995), where an essential component of the model is the 
Compton thick shielding gas between the X-ray and UV emission, protecting the disk wind from being ionized by the X-ray emission. 
\citet{b21} also reached a similar conclusion, where the Thomson scattering optical depth was argued to have $\tau \ge 3$ 
($\ge 20$ times dimmer). In this paper, the constraint is likely more stringent with $\tau \sim 6$ ($\sim 400$ times dimmer). Aoki 
(2010) analyzed the UV spectrum of J1723+5553, and found the \nh\ column density in the UV wind was $\nh \ge 5\times10^{17}$ \cmsq\
 using the curve of growth method with the unresolved Balmer absorption lines. Aoki (2010) pointed out 
that the covering fraction cannot be determined due to unresolved absorption lines, and assumed a covering fraction of 1. The column 
density would be higher than the given limit if the covering fraction is significantly less than 1. The limit derived by 
Aoki (2010) is seven orders of magnitude less than, but still consistent with, our X-ray column densities. If the X-ray and UV 
absorbing gasses are located in different regions, it can still be incorporated into both existing geometrical interpretations of FeLoBALs 
(e.g., Murray et al.\ 1995; Gallagher \& Everett 2007) or evolutionary models (e.g., Fabian 1999). However, for evolutionary models 
the UV emission should come from the photosphere of the gas/dust cloud rather than the accretion disk. Such models have been 
simulated in recent studies \citep[e.g.,][]{ca08}, and support that FeLoBALs may be an evolutionary stage in the development of normal quasars.

Another explanation is that FeLoBALs have intrinsic SEDs different from normal quasars. They could be extremely X-ray weak compared to normal quasars, as 
in the case of the narrow-line quasar PHL 1811 \citep{l07}, or the Narrow-Line Seyfert 1 galaxy WPVS 007 \citep{g08}. This would nullify our \nh\ 
constraints for J0934+5417 and J1352+4239, since they are obtained by assuming a normal quasar SED. However, this would not explain J1723+5553, which we 
detect in the observed 2--10~keV band, but not in the observed 0.2-2~keV band. If J1723+5553 were intrinsically X-ray weak, and X-ray unabsorbed, we would 
expect to see a flat-line power-law slope for quasars in the hard and the soft X-ray bands. This is not consistent with our observations. In addition, the \nh\ 
constraint from the X-ray hardness ratio is still valid for $\nh \ge 6\times10^{23}$ \cmsq\ with the Thomson scattering optical depth $\tau \ge 0.4$, which could 
only marginally scatter the UV emission if the true \nh\ column density is close to the lower boundary. 

More X-ray detections of FeLoBALs are needed so that we are not restricted by using the broadband SED constraints. It is also 
possible that quasar host galaxies contribute a 
fraction of rest frame 4000--7000\AA\ emission; however, the contamination at this band is usually small. In addition, we correct the intrinsic 
dust extinction for FeLoBALs using the average dust extinction for LoBALs, which will underestimate the intrinsic flux and give 
less steep \aox\ values. A priori, it is 
unclear what the net effect of these two possibilities will be on the broadband flux; IR spectroscopy is needed to help resolve these issues.

\subsection{AGN Kinetic Feedback}
We calculate the kinetic feedback efficiency, $\epsilon_k = \dot{E}_k/L_{Bol}$, of the X-ray absorber in FeLoBALs using the location, 
luminosity, and \nh\ column density constraints obtained in this paper. In particular, $\dot{E}_k$ is the kinetic feedback 
power, $\dot{E}_k = \dot{M}v^2/2 = 4\pi\mu m_p f_c r \nh v^3$ (e.g., Moe et al.\ 2009), where $\mu$ is the mean molecular weight, 
$m_p$ is the proton mass, $f_c$, $r$, \nh\ and $v$ are the covering fraction, location, column density, and velocity of the wind,
 respectively, and $L_{Bol}$ is the bolometric luminosity of the quasar.  If the SEDs of FeLoBALs are consistent with those of 
normal quasars, we obtain a reasonable $\nh \sim 3\times10^{24}$ \cmsq; this is the upper end of the column density 
range for our 3$\sigma$ detection, and entirely consistent with our column density lower limits for the non-detections. We locate 
the wind between the UV and X-ray emission regions, $\sim 40 r_g$ and $r_g = GM/c^2$, using the microlensing constraints of 
Dai et al.\ (2010a). For the covering fraction, we use the intrinsic FeLoBAL population 
range of 1.5\% and 2.1\%, depending on which model \citep{b12} is used, 
and assume the quasars are emitting at a typical 0.3 Eddington limit (Shankar et al.\ 2010). We are left with one uncertainty, 
the velocity of the X-ray absorber. The velocity of the BAL wind is usually measured in the UV wind, which can reach up to 
$\sim 0.1c$. Since the X-ray absorbing wind is located at smaller radii than the UV absorber, its velocity can be higher than the 
wind velocity measured from the UV spectrum. The only velocity measurement of the X-ray absorber is from the blue-shifted X-ray 
absorption lines detected in a few gravitationally lensed BALQSOs (Chartas et al.\ 2002, 2003, 2007), and this velocity can reach 
0.3--0.8$c$. Since the X-ray absorption lines are mostly detected in mini-BALs, we could possibly be looking through the edge of the wind, where 
the wind can be by fully accelerated. Thus, we consider the X-ray absorption line as providing an upper limit, and assume our 
wind velocity is 0.1--0.3$c$, between the estimates from the two methods. We find the feedback efficiency, $\epsilon_k$, for FeLoBALs 
is in the range (outflow velocity dependent) of 
either 0.2\%--4.8\% for a covering fraction of 1.5\%, or 0.3\%--6.9\% for a covering fraction of 2.1\%. To reach the minimum of 
5\% required to explain the co-evolution between black holes and host galaxies (e.g., Silk \& Rees 
1998; Granato et al.\ 2004; Hopkins et al.\ 2005), with the other variables given here, we need a column density of 
\nh$\ge8\times10^{25}$ \cmsq\ or \nh$\ge6\times10^{25}$ \cmsq\ for low velocity wind with the different covering fractions. These are consistent 
with our lower limits on column density for the two non-detections, but out of reach for J1723+5553. However, for the high end of the conservative 
wind velocity range, only \nh$\ge3\times10^{24}$ and \nh$\ge2\times10^{24}$ are needed for $f_c=$1.5\% and 2.1\%, respectively. Both of these 
column densities are consistent with the results for J1723+5553, and thus
it is likely that FeLoBALs can contribute to AGN feedback. 
This becomes even more likely when models like \citet{he10} are considered. They describe a \textquotedblleft two-stage\textquotedblright\ feedback process, 
where a weak wind from the central engine energizes a hot, diffuse interstellar medium, which is then amplified when it it hits 
instabilities in a cold cloud within the host. This amplification outside of the central engine requires feedback efficiency as small as 0.5\%. The 
BAL wind in FeLoBALs, therefore, is a promising candidate for the feedback 
process responsible for the co-evolution between black holes and host galaxies. More measurements of $N_H$ are needed to further quantify 
how important the contributions are from FeLoBALs and BALQSOs to kinetic feedback efficiency.

\acknowledgements 
This research has made use of data obtained from the Suzaku satellite, a collaborative mission between the space agencies of Japan (JAXA) and the USA (NASA).  
We acknowledge the financial support by NASA grant NNX09AV68G. FS acknowledges the support from the Alexander von Humboldt Foundation.

Funding for the SDSS and SDSS-II has been provided by the Alfred P. Sloan Foundation, the Participating Institutions, the National Science Foundation, 
the U.S. Department of Energy, the National Aeronautics and Space Administration, the Japanese Monbukagakusho, the Max Planck Society, and the 
Higher Education Funding Council for England. The SDSS Web Site is http://www.sdss.org/.

\clearpage

\end{document}